\documentstyle[aps,twocolumn]{revtex}
 
\begin{document}
 
{\bf Comment on ``Quasiparticle
  Spectra around a Single Vortex in a $d$-wave Superconductor''}

In a recent Letter\cite{mkm} Morita, Kohmoto and Maki (MKM) analyzed 
the structure of quasiparticle states around a single vortex in a 
$d_{x^2-y^2}$ superconductor using an approximate version of 
Bogoliubov - de Gennes (BdG) theory with a model $d$-wave gap function.
Their principal result is the existence of a bound state within the core
region
at finite energy with full rotational symmetry, which they assert explains
the recent scanning tunneling microscopy
 results on YBa$_2$Cu$_3$O$_{7-\delta}$ single crystals \cite{maggio}. 
Here we argue that the approximate BdG equations used by MKM are
fundamentally inadequate for the description of a $d$-wave vortex and that
the obtained circular symmetry of the local density of states (LDOS) is an
unphysical artifact of this approximation. 

BdG equations for a $d$-wave superconductor differ from the conventional
ones by the non-locality of the off-diagonal term $\int d{\bf x}'
\tilde\Delta({\bf x},{\bf x}')v({\bf x}')$.  MKM approximate this term by
a local term $\Delta({\bf k};{\bf R})v({\bf R})$, 
and claim that this is accurate to
$O(1/k_F\xi)$, quoting the derivation given by Bruder\cite{bruder}. A key
step in Bruder's calculation is factoring out the rapid $k_F$ oscillation
in $v({\bf x})$ by introducing a slowly varying envelope function
$\bar{v}_{\bf k}({\bf x})$: 
\begin{equation} v({\bf x}) =
e^{ik_F{\bf\hat k}\cdot{\bf x}} \bar{v}_{\bf k}({\bf x}). 
\end{equation}
$v({\bf x})$ is then substituted in the off-diagonal matrix element and
after transforming to the center of mass frame both $\Delta$ and $\bar{v}$
are Taylor-expanded in the relative coordinate. The leading term in the
expansion is that used by MKM and corrections are indeed formally
$O(1/k_F\xi)$. Since one cannot expand the rapidly varying function
$v({\bf x})$ directly, this manipulation rests heavily on the ansatz (1).
An implicit assumption made in (1) is that the wave-vector ${\bf k}$ is a
good quantum number. This is approximately correct in the simple planar
scattering situations considered by Bruder\cite{bruder}. However, ${\bf
k}$ is certainly {\em not} a good quantum number near the vortex core; the
appropriate ansatz in this situation would involve a sum over ${\bf \hat
k}$ in Eq.\ (1). One then obtains the leading off-diagonal term of the
form $\sum_{\bf{\hat k}} e^{ik_F\bf{\hat k}\cdot\bf{R}}\Delta({\bf k};{\bf
R})\bar v_{\bf k}({\bf R})$, indicating that all the plane wave states are
mixed together by the non-trivial structure of the gap. This in turn
results in coupling of the various angular momentum channels and the loss
of rotational symmetry. 
MKM improperly neglect the summation over ${\bf \hat k}$ in the ansatz 
(1) for the wave functions. 

Inadequacy of MKM's approximation becomes even more transparent when
analyzing its implications for the physical quantities, such as the LDOS.
A brief examination of the (standard) BdG equations for a $d$-wave vortex
reveals that both the quasiparticle amplitudes and the self-consistent gap
function will necessarily exhibit a four-fold anisotropy. This qualitative
conclusion is corroborated by the numerical results within the
Eilenberger\cite{ichioka} and the lattice BdG\cite{wang} formalisms.
Anisotropic wave-functions are also found near a strongly scattering
non-magnetic impurity\cite{balatsky}. These anisotropies are the salient
feature of unconventional superconductivity resulting from an intricate
interplay between the center of mass and the relative coordinate in the
off-diagonal part of the BdG Hamiltonian. The fact that such anisotropies
have not yet been observed experimentally\cite{maggio} is indeed puzzling
and apparently inconsistent with the above theoretical work. However,
reconciliation between the theory and the experiment cannot be brought
about by simply ignoring the symmetry breaking interactions as is done by
MKM. 

A brief inspection of MKM Eqs.\ (1) and (2) reveals that they in fact
describe a set of decoupled $s$-wave vortices with the gap functions of
various amplitudes (but identical profiles) scaled by the factor
$\cos[2\theta({\bf\hat k})]$.  Each of these vortices possesses a set of
bound states\cite{caroli} with the energy spacing $\Delta^2({\bf
k};\infty)/E_F$, contributing sharp peaks in the LDOS. The finding of MKM
of the broad peak in the LDOS in the vicinity of the core is a trivial
consequence of superimposing the peaks corresponding to the lowest bound
states belonging to these $s$-wave vortices.  
The full rotational symmetry of the LDOS in MKM's result also directly 
follows from this observation. Clearly, such calculation
does not resolve the interesting question of the existence and the
structure of the quasiparticle bound states in $d$-wave vortices. 
We expect that the nature of such states will be qualitatively
different from the conventional picture of Caroli, de Gennes and Matricon
\cite{caroli}. However, this structure can emerge only from a solution, 
numerical or analytical, which preserves complexity and richness
of the original BdG theory.


\vspace*{10pt}

\noindent 
 M. Franz$^1$ and M. Ichioka$^2$ 
\\ \hspace*{5pt}
\begin{minipage}[t]{3in} $^1$Department of Physics and Astronomy
\\ ~\ Johns Hopkins University
\\ ~\ Baltimore, MD 21218
\\ $^2$Department of Physics, Okayama University
\\ ~\ Okayama 700, Japan
\end{minipage}

\vspace*{10pt}

\noindent \today 

\end{document}